# Distilling High Diagnostic Value Patches for Whole Slide Image Classification Using Attention Mechanism

Tianhang Nan[1], Hao Quan[1], Yong Ding[1], Zhir Zhang[1], Weij Xu[1], Xiaoyu Cui[1, 2], Hang Lu[3]
[1]The College of Medicine and Biological Information Engineering, Northeastern University, Shenyang, China.
[2]The Key Laboratory of Biomedical Imaging Science and System, Chinese Academy of Sciences
[3]Jinzhou Medical University, Jinzhou, China.

**Abstract**—Multiple Instance Learning (MIL) has garnered widespread attention in the field of Whole Slide Image (WSI) classification as it replaces pixel-level manual annotation with diagnostic reports as labels, significantly reducing labor costs. Recent research has shown that bag-level MIL methods often yield better results because they can consider all patches of the WSI as a whole. However, a drawback of such methods is the incorporation of more redundant patches, leading to interference. To extract patches with high diagnostic value while excluding interfering patches to address this issue, we developed an attention-based feature distillation multi-instance learning (AFD-MIL) approach. This approach proposed the exclusion of redundant patches as a preprocessing operation in weakly supervised learning, directly mitigating interference from extensive noise. It also pioneers the use of attention mechanisms to distill features with high diagnostic value, as opposed to the traditional practice of indiscriminately and forcibly integrating all patches. Additionally, we introduced global loss optimization to finely control the feature distillation module. AFD-MIL is orthogonal to many existing MIL methods, leading to consistent performance improvements. This approach has surpassed the current state-of-the-art method, achieving 91.47% ACC (accuracy) and 94.29% AUC (area under the curve) on the Camelyon16 (Camelyon Challenge 2016, breast cancer), while 93.33% ACC and 98.17% AUC on the TCGA-NSCLC (The Cancer Genome Atlas Program: non-small cell lung cancer). Different feature distillation methods were used for the two datasets, tailored to the specific diseases, thereby improving performance and interpretability.

## 1. INTRODUCTION

Pathological diagnosis serves as crucial preoperative support for tumor treatment, traditionally conducted by pathologists using microscopes [1]. As digital slide scanning technology has become increasingly reliable, pathological slides are saved as whole slide images (WSIs) and stored and viewed on computers for clinical and research purposes [2]. Subsequently, deep learning has also been applied in assisting pathological image diagnosis [3-5]. Given the large size of WSI images and the inherently high risks associated with medical tasks, computational pathology [1] poses a significant challenge in the field of computer vision. In this domain, deep learning techniques are employed for tasks such as WSI classification [1, 6-8] and segmentation [9-11], contributing [12-21] to precise disease diagnosis, prognosis and treatment. Existing models in the field of computer vision are predominantly designed for smaller images (typically, 256×256 pixels or the comparable magnitude). When dealing with WSIs, which often have gigapixel dimensions, the common practice is to divide them into multiple patches. However, in tasks such as tumor diagnosis, labeling patches at the patch level or annotating tumor regions at the pixel level involves prohibitively high manual labor costs [22].

This challenge has prompted researchers to explore weakly supervised solutions for addressing issues in the WSI classification field. MIL has achieved significant success [1][6] because it can be trained using only the reported diagnostic results as labels and can provide prediction probabilities for each region within the WSI, thus proving to be successful in assisting diagnoses. However, several challenges still remain [23-26]. For instance, due to the large number of patches present in WSIs, the training process can introduce a significant amount of redundant features, which may interfere with the model's learning process. In addition, the inference criterion of instance-level MIL is that when any patch within a WSI is diagnosed as malignant, the entire WSI is classified as malignant. While this rule may seem reasonable, in real-world scenarios, one false positive patch can lead to the misclassification of a WSI as malignant, significantly reducing precision.

In recent years, there has been a growing focus among researchers on bag-level MIL methods [8, 27, 28]. These approaches extract features from all patches within a WSI and further fuse them into a single bag-level feature to predict the class of the WSI. These methods have proven more effective in pathology image classification tasks. However, most research in this direction has focused on improving the feature fusion module to obtain better bag-level features. For example, from feature score-weighted fusion to the development of fusion based on self-attention, bag-level features have been better represented, leading to continuous improvement in model performance. However, an important issue has been overlooked: the number of high-diagnostic-value patches (such as substantive tumors) in WSIs is significantly lower than the number of interfering patches (such as non-cellular tissue and benign cells). Merely improving the feature fusion module is insufficient to eliminate the negative impact of redundant original inputs on model performance.

To distill high-diagnostic-value subregions within WSIs and exclude most interference before feature fusion, we designed an attention-based feature distillation multi-instance learning (AFD-MIL) approach, as shown in Figure 1. The first step involves constructing dual-channel feature distillation modules based on attention mechanisms and weakly supervised instance-level classifiers. These modules are designed to distill the features from all original patches, eliminate redundant feature interference, and emphasize more valuable features. In the second step, the distilled features are further transformed into WSI-level features through feature fusion. These features are then used for classification tasks. Finally, a global loss optimization is introduced when designing the loss function, directly adjusting the loss optimization of the feature distillation module based on the quality of the final classification results. Most existing MIL methods are orthogonal to AFD-MIL, allowing seamless integration into existing solutions, thus enhancing performance and contributing to further research in the field. In the first section of results, AFD-MIL was compared with the latest baselines on the Camelyon16 [29] and TCGA-NSCLC [30] datasets, and consistently achieved SOTA performance. Secondly, the impacts of various model components on overall performance were investigated, including feature distillation scale, ablation experiments, and different distillation methods. Different distillation methods were identified to adapt to different tasks (including benign-malignant classification and tumor subtypes classification). Finally, we visualized the regions of interest in AFD-MIL, demonstrating its potential for tumor detection. The code is available at https://github.com/MasyerN/AFD-MIL. In summary, the main contributions of this study are as follows:

(1) This work introduces the use of dual-channel feature distillation as a preliminary control for selecting high-diagnostic subregions in Whole Slide Images (WSIs), directly relieving the challenge posed by excessive redundancy in features.
(2) This study introduces the use of attention mechanisms to distill features with high diagnostic value, as opposed to the traditional approach of indiscriminately and forcibly fusing all patches, thereby achieving optimal performance in cancer classification tasks (Camelyon16 and NSCLC).
(3) AFD-MIL achieves high interpretability as it automatically selects regions similar to ground truth. Additionally, its compatibility with various existing models enhances performance, demonstrating high scalability.

## 2. Related Work

### 2.1 Instance- level Multi-Instance Learning

Instance-level MIL [6, 7, 23-26, 31] methods are trained by optimizing the features of individual patches. During the training phase of these methods, a filter is applied to select the "top k" patches with the highest malignant probability. Subsequently, the training process involves backpropagation with WSI-level labels assigned to the selected patches. This process enables the encoder and fully connected layers responsible for patch classification to learn to differentiate between benign and malignant tissues. During inference, the model classifies each patch to determine its category, and when any patch within a WSI is identified as "malignant", the entire WSI is classified as malignant. However, such approaches tend to yield logical errors when faced with multi-diseases classification tasks such as the NSCLC dataset [30]. Because in this dataset, positive and negative samples represent two subtypes of cancer, rather than "malignant" and "benign". Multiple studies have demonstrated that instance-level MIL performs less effectively than do bag-level MIL methods [8][27].

### 2.2 Bag-level Multi-Instance Learning

Bag-level MIL methods typically employ feature fusion techniques to combine the features of all instances into global features at the WSI level [8, 27, 28, 31-35]. Common approaches involve utilizing feature weighting [8] or self-attention mechanisms [32]. To reduce the interference of redundant features during feature fusion, researchers have developed two-step multi-instance learning approaches [33]. In this approach, prior to feature fusion, probabilities indicating the presence of tumors in each patch are obtained

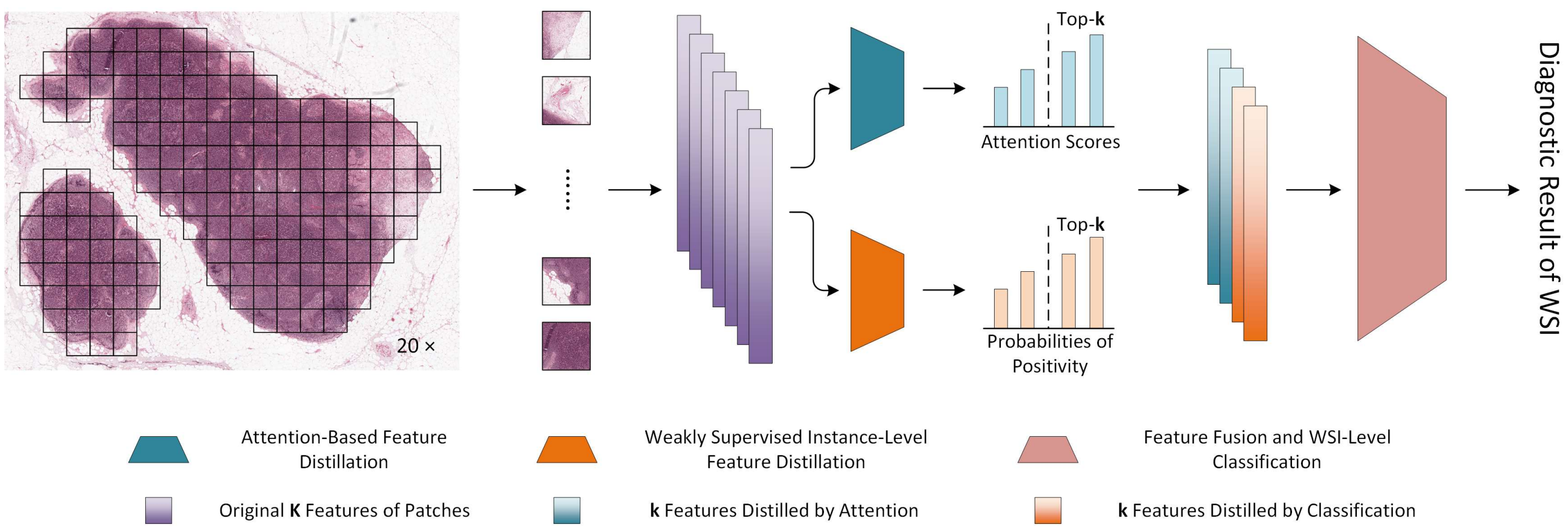

**Fig. 1.** The AFD-MIL model utilizes a feature distillation module to extract effective patch-level features. The features are fused and a classification network is utilized to obtain the WSI-level classification results. The network requires only WSI-level labels. Detailed annotations for each patch are not required.

**Algorithm 1:** Weakly Supervised Instance-level Feature Distillation

**Input:** instance features $X = \{x_1, x_2, \dots, x_K\}$ of WSI $W$, the label $Y$ of $W$, distilled feature number $k$

**Output:** distilled features $\{h_1, h_2, \dots, h_k\}$, loss function $L_1$

$\{\hat{y}_1, \hat{y}_2, \dots, \hat{y}_K\} \leftarrow 0$
**for** $1 \le i \le K$ **do**
  $\hat{y}_i \leftarrow MLP_1(x_i)$
  $i \leftarrow i + 1$
**end for**
$\{\hat{h}_1, \hat{h}_2, \dots \hat{h}_k\} \leftarrow argmax((\hat{y}_1, \hat{y}_2, \dots \hat{y}_K), k)$
$L_1 \leftarrow 0$
**for** $h \in \{\hat{h}_1, \hat{h}_2, \dots \hat{h}_k\}$, $1 \le i \le k$ **do**
  $h_i \leftarrow x_{argindex(h)}$
  $L_1 \leftarrow L_1 + crossentropyloss(\hat{h}_1, Y)$
  $i \leftarrow i + 1$
**end for**
**return** $\{h_1, h_2, \dots, h_k\}$, $L_1$

through instance-based multi-instance learning. These probabilities are then used to select the desired features for further fusion. However, such methods do not fully address the inherent issues of instance-level MIL methods (the first step), including their suboptimal performance during feature selection and limited interpretability. Additionally, the subsequent network training is decoupled, making it challenging to fine-tune the network based on the final classification results. In this study, the problems associated with two-step multi-instance learning are discussed and optimized, leading to further improvements in model performance.

## 3. Method

The model used in this study is presented in Figure 1. First, a WSI is segmented into several patches, each sized appropriately for neural network processing (the size of each patch is 256×256 pixels, and a WSI at the magnification of 20× is usually divided into over 10,000 patches). Then, these patches are encoded into features using a pretrained image encoder. The dual-channel feature distillation model refines these features for feature fusion, ultimately making predictions. The detailed structure of the model is depicted in Figure 2, which shows two distinct feature distillation models and the subsequent classification model.

This section is divided into two subsections for discussion. Subsection 1 introduces the dual-channel feature distillation model, which corresponds to Algorithms 1 and 2. Subsection 2 presents information on feature fusion, WSI-level classification, and global loss optimization, which constitute the Algorithm 3. Algorithm 3 is still a comprehensive summary of the AFD-MIL model.

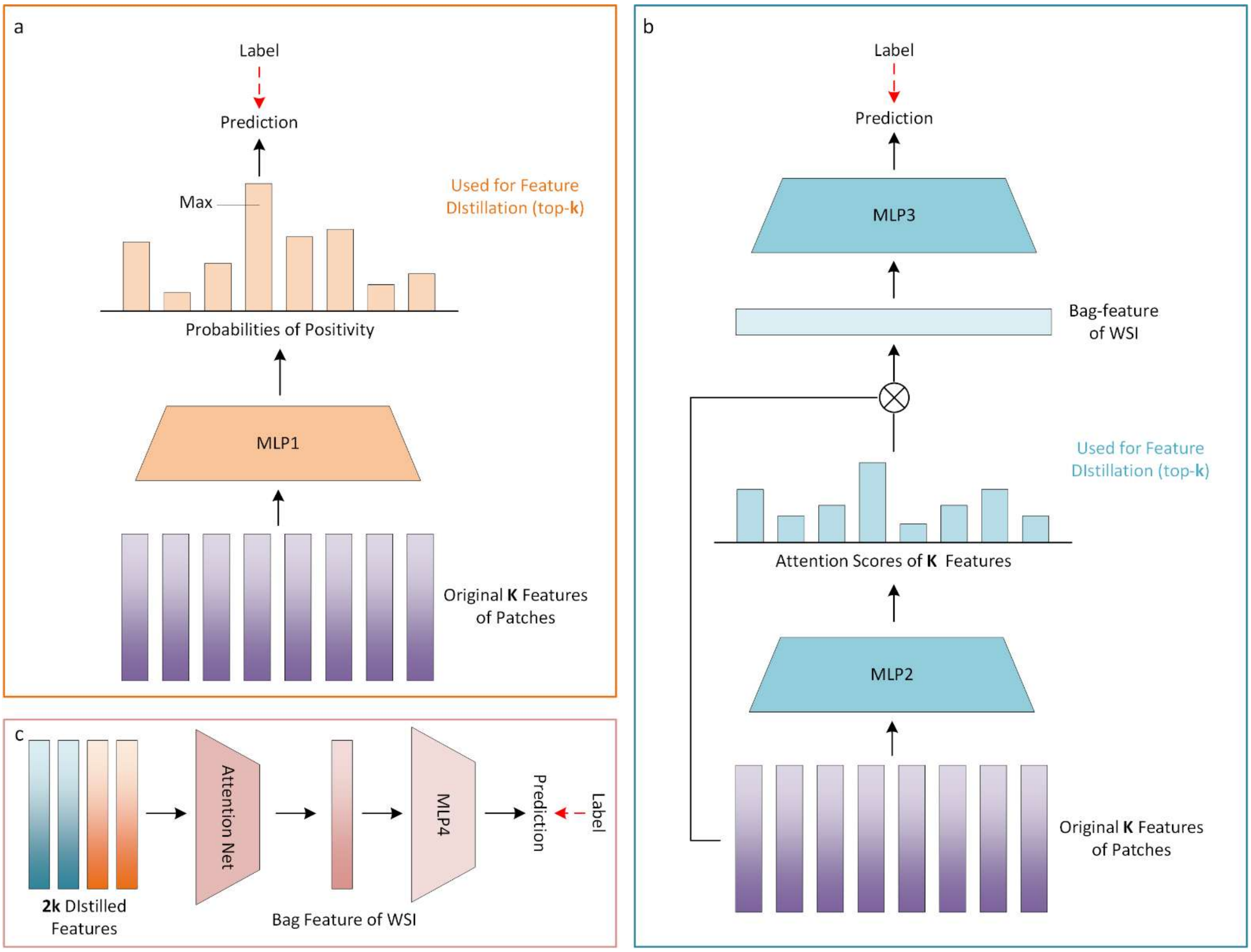

**Fig. 2.** The detailed structure of the feature distillation modules and the classification module. "MLP" is an abbreviation for Multilayer Perceptron. **a.** Weakly Supervised Instance-level Feature Distillation; **b.** Attention-based feature distillation; and **c.** Feature fusion and WSI-level classification.

### 3.1 Feature Distillation

1) Weakly Supervised Instance-level Feature Distillation

For a given WSI $W$, patches can be obtained through tissue region segmentation [28] and dividing to 256×256-pixel size. Then, a pretrained image encoder is utilized to extract features from these patches. This model is not involved in the training process. The model designed in this study focuses solely on learning from the extracted features. As shown in Fig. 2.a and Algorithm 1, the patches yield instance-level features $X = \{x_1, x_2, \ldots, x_K\}$, where $K$ represents the total number of patches contained in $W$. Each individual patch $x_i$ has a latent label $y_i$ (where $y_i = 1$ denotes positive, and $y_i = 0$ denotes negative). These labels exist in reality but are unknown to the model. The labels $Y$ for WSI $W$ are known to the model and are represented as:

$$Y = \begin{cases} 1, if \sum_{i=1}^{K} y_i > 0 \\ 0, if \sum_{i=1}^{K} y_i = 0 \end{cases} \tag{1}$$

When at least one patch in WSI $W$ is classified as positive, the WSI is categorized as positive; otherwise, the WSI is classified as negative. The patch classification network $f_{cla}$ typically consists of fully connected layers designed for the patch classification task: $\hat{y}_i = f_{ins}(x_i)$, where $\hat{y}_i$ represents the probability of $x_i$ being predicted as positive. Once predictions for all patches in $W$ are made, the labels $Y$ are used to calculate the loss by considering the top-$k$ highest predicted probabilities $\{\hat{h}_1, \hat{h}_2, \ldots \hat{h}_k\}$from $\{\hat{y}_1, \hat{y}_2, \ldots, \hat{y}_K\}$. The gradient is then backpropagated accordingly, enabling the weakly supervised learning of $f_{ins}$:

$$\hat{h}_1, \hat{h}_2, \ldots \hat{h}_k = argmax((\hat{y}_1, \hat{y}_2, \ldots \hat{y}_K), k) \tag{2}$$

**Algorithm 2:** Attention-Based Feature Distillation

**Input:** Instance features $X = \{x_1, x_2, \dots, x_K\}$ of WSI $W$, the label $Y$ of $W$, distilled feature number $k$

**Output:** Distilled features $\{o_1, o_2, \dots, o_k\}$, loss function $L_2$

$\{\alpha_1, \alpha_2, \dots, \alpha_K\} \leftarrow softmax\ (MLP_2(X))$
$feature_{atten} \leftarrow \vec{0} \in R^{n \times 1}$
**for** $1 \le i \le K$ **do**
$feature_{atten} \leftarrow feature_{atten} + \alpha_i * x_i$
$i \leftarrow i + 1$
**end for**
$\hat{Y} \leftarrow MLP_3(feature_{atten})$
$L_2 \leftarrow crossentropyloss(\hat{Y}, Y)$
$\{\hat{o}_1, \hat{o}_2, \dots \hat{o}_k\} \leftarrow argmax((\alpha_1, \alpha_2, \dots, \alpha_K), k)$
**for** $o \in \{\hat{o}_1, \hat{o}_2, \dots \hat{o}_k\}$, $1 \le i \le k$ **do**
$o_i \leftarrow x_{argindex(o)}$
$i \leftarrow i + 1$
**end for**
**return** $\{o_1, o_2, \dots, o_k\}$, $L_2$

**Algorithm 3:** Attention-Based Feature Distillation MIL

**Input:** Instance features $X = \{x_1, x_2, \dots, x_K\}$ of WSI $W$, the label $Y$ of $W$, distilled feature number $k$

**Output:** Predicted classification result $Y_{pred}$ of $W$

Initialize model parameters
$\{h_1, \dots, h_k\}$, $L_1$ $\leftarrow$ **Algorithm 1**$(X,\ Y,\ k)$
$\{o_1, \dots, o_k\}$, $L_2$ $\leftarrow$ **Algorithm 2**$(X,\ Y,\ k)$
$\{feature_1, feature_2, \dots, feature_{2k}\} \leftarrow \{h_1, \dots, h_k, o_1, \dots, o_k\}$
$feature_W \leftarrow f_{fusion}(\{feature_1, feature_2, \dots, feature_{2k}\})$
$\hat{Y}_{final} \leftarrow MLP_4(feature_W)$
$Y_{pred} \leftarrow argmax(\hat{Y}_{final})$
$L_3 \leftarrow crossentropyloss(\hat{Y}_{final}, Y)$
$L \leftarrow (loss_1 + loss_2) * exp(-||loss_3||_1) + loss_3$
Update parameters of all models (Algorithm 1, Algorithm 2, $MLP_4$)
**return** $Y_{pred}$

$$loss_1 = -\frac{1}{k}\sum_{i=0}^{k} Y * log(\hat{h}_i) + (1 - Y) * log(1 - \hat{h}_i) \tag{3}$$

The top-$k$ features with highest predicted probabilities selected by $f_{ins}$, denoted as $\{h_1, h_2, \dots, h_k\}$, are utilized for subsequent feature fusion.

During this feature distillation process, patches with higher positive probabilities are selected. This approach is more effective for the tumor versus normal tissue classification task with the Camelyon16 dataset. However, this approach is not as effective for the NSCLC dataset. The reason for this difference lies in the nature of the NSCLC dataset, where both positive and negative samples represent various types of tumors, and there are other regions in WSIs that can be considered “normal tissue”. The patches in these areas fall into a “third category” of images that are neither negative nor positive. Defining such images as negative samples would be incorrect. Therefore, we compare another feature distillation approach, i.e, an approach where the top $k/2$ features with the highest positive probabilities and the top $k/2$ features with the highest negative probabilities are selected. These features are collectively used for subsequent feature fusion and WSI classification tasks. These two feature distillation methods are referred to as **Max-Positive** and **Max-Positive& Negative**, respectively.

2) Attention-Based Feature Distillation

As shown in Fig. 2.b and Algorithm 2, the role of the attention network $f_{atten}$ is to obtain attention scores $\alpha_i$ corresponding to the instance $x_i$. Then, by weighting the instance features, the WSI-level global feature $X_{wsi}$ is generated. Subsequently, a fully connected layer is used to predict the labels for the WSI:

$$\{\alpha_i\}_1^K \in \mathbb{R}^{K\times 1} = f_{atten}(\{x_i\}_1^K \in \mathbb{R}^{K\times n}) \tag{4}$$

$$X_{wsi} = \sum_{i=1}^{K} \alpha_i * x_i \tag{5}$$

$$\hat{Y} = MLP(X_{wsi}) \tag{6}$$

The results obtained by $f_{atten}$ during the feature distillation phase are trained using cross-entropy loss, but WSI-level predictions are not obtained during the inference stage. These results are solely used for feature distillation. $f_{atten}$ selects the top-$k$ features with the highest attention scores $\{\hat{o}_1, \hat{o}_2, \dots \hat{o}_k\}$ from $\{\alpha_1, \alpha_2, \dots \alpha_K\}$, denoted as $\{o_1, o_2, \dots, o_k\}$, for subsequent feature fusion:

$$loss_2 = Y * log(\hat{Y}) + (1 - Y) * log(1 - \hat{Y}) \tag{7}$$

$$\hat{o}_1, \hat{o}_2, \dots \hat{o}_k = argmax((\alpha_1, \alpha_2, \dots \alpha_K), k) \tag{8}$$

3.2 Feature Fusion, WSI-Level Classification and Global Loss

A total of $2k$ distilled features from $W$ are obtained through $f_{ins}$ and $f_{atten}$. The methods for feature fusion are diverse; our AFD-MIL is not restricted to any specific feature aggregator, including the architecture and training paradigms. Please refer to Section 4.2 for details on our selections. As shown in Figure 2.c, the fused features are then used for the classification of WSIs according to the following formulas:

$$\hat{Y}_{final} = MLP[f_{fusion}(h_1, h_2, \dots h_k, o_1, o_2, \dots o_k)] \tag{9}$$

$$loss_3 = Y * log(\hat{Y}_{final}) + (1 - Y) * log(1 - \hat{Y}_{final}) \tag{10}$$

The traditional two-step MIL feature distillation optimization results are not related to the final WSI classification results and cannot validate the impact of feature distillation parameters on the ultimate classification results. To address this issue, we introduce global loss function optimization. In the final loss function, we utilize the WSI predictions to optimize the loss function for the feature distillation section:

$$loss = (loss_1 + loss_2) * exp(-||loss_3||_1) + loss_3 \tag{11}$$

This loss function can be optimized for feature distillation based on the value of $loss_3$, which represents the model's final prediction. Adjusting the gradient backpropagation during training is achieved by applying additional weights to $loss_1$ and $loss_2$. Favorable final predictions will provide incentive for the feature distillation step in the current batch, while unfavorable predictions will be penalized.

## 4. RESULTS

4.1 Datasets and Data Preprocessing

Two publicly available datasets were used for model training and evaluation: Camelyon16 [29] and TCGA-NSCLC [30]. Camelyon16 is a histopathology image dataset for breast cancer metastasis detection. This dataset consists of 399 WSIs. WSIs without background were divided into 256×256-sized patches at 20× magnification. There are approximately 2.8 million patches in total. TCGA-NSCLC includes two subtypes of lung cancer: lung adenocarcinoma and lung squamous cell carcinoma. This dataset comprises 1054 WSIs, which were further divided into approximately 5.2 million patches at 20× magnification. These patches were encoded into feature vectors using the ResNet18 [31] pretrained by ImageNet [32]. The encoder was not involved in model training, which is common in the MIL research series. It is worth noting that the use of advanced encoders or fine-tuning the training dataset to improve model performance is widely acknowledged. However, enhancing performance in this manner is beyond the scope of discussion. Like previous studies [34-36], we also use the same low-performance image encoder to highlight the superiority of our weakly supervised learning framework.

TABLE 1

DETAILS OF THE DATASETS

| Item | Training | | Testing | |
|---|---|---|---|---|
| | Negative | Positive | Negative | Positive |
| WSIs in Camelyon16 | 154 | 111 | 80 | 49 |
| Patches in Camelyon16 | 1437600 | 1036187 | 737829 | 451919 |
| WSIs in TCGA-NSCLC | 427 | 409 | 107 | 103 |
| Patches in TCGA-NSCLC | 1683444 | 1524389 | 440289 | 423794 |

TABLE 2

WSI CLASSIFICATION RESULTS

| Methods | Camelyon16 | | | | TCGA-NSCLC | | | |
|---|---|---|---|---|---|---|---|---|
| | ACC (%) | AUC (%) | Recall (%) | Precision (%) | ACC (%) | AUC (%) | Recall (%) | Precision (%) |
| CLAM-SB | 86.05 | 86.82 | 73.47 | 87.80 | 89.52 | 89.36 | 80.58 | **97.65** |
| CLAM-MB | 86.82 | 85.10 | 69.39 | **94.44** | 86.19 | 89.59 | **99.03** | 78.46 |
| ABMIL | 84.50 | 84.07 | 81.71 | 86.71 | 81.43 | 88.95 | 85.84 | 82.75 |
| TransMIL | 83.72 | 81.29 | 81.06 | 85.43 | 85.24 | 90.70 | 85.31 | 85.46 |
| DTFD-MaxS | 82.95 | 82.77 | 80.09 | 84.85 | 81.9 | 88.91 | 83.77 | 82.29 |
| IB-DTFD-MaxS | 88.37 | 89.51 | 86.51 | 89.53 | 82.86 | 90.5 | 82.96 | 83.25 |
| IB-ABMIL | 88.37 | 90.43 | 87.14 | 88.58 | 85.24 | 91.26 | 85.17 | 85.42 |
| IB-TransMIL | 83.72 | 88.71 | 82.93 | 83.14 | 85.24 | 92.54 | 87.06 | 85.80 |
| AFD-ABMIL | 90.70 ↑6.20 | 90.52 ↑6.45 | **89.80** ↑8.09 | 86.27↓0.44 | **93.33** ↑11.9 | **98.17** ↑9.22 | 93.26 ↑7.95 | 93.64 ↑10.89 |
| AFD-TransMIL | **91.47** ↑7.75 | **94.29** ↑13.0 | 85.71 ↑4.65 | 91.30 ↑5.87 | **93.33** ↑8.09 | 97.98 ↑7.28 | 94.17 ↑8.86 | 92.38 ↑6.92 |

For the training and testing data split approach and the image encoding process, the guidelines outlined in Reference [34-36] were followed. This data partitioning method ensures a balanced representation of samples from different classes. Moreover, it is widely accepted, making comparisons with baselines more intuitive and reproducible. The data partition is depicted in Table 1 (consistent with the method used in the abovementioned study, some corrupted WSIs were removed).

### 4.2 Baselines and Evaluation Metrics

Eight bag-level MIL algorithms were selected for comparison. Among them, we built AFD-MIL on the basis of two classical algorithms, TransMIL (Transformer MIL) [34] and ABMIL (Attention Based MIL) [8]. The other six algorithms are CLAM-SB [28], CLAM-MB [28], DTFD-MaxS [35], IBMIL-DTFD [36], IBMIL-ABMIL [36], and IBMIL-TransMIL [36] (IBMIL models are SOTA). All baselines were configured according to their official settings where available. Additionally, our replication results for the baselines are the same or similar to those in [34-36], which are convincing.

For the WSI classification task, the accuracy (ACC) and area under the receiver operating characteristic (ROC) curve (AUC) are the most important evaluation metrics. Given the high-risk nature of medical tasks and the fact that existing assistive diagnostic solutions are primarily used to help pathologists exclude negative samples, recall and precision are also important evaluation metrics. In comparison to the classification task of two cancer subtypes in NSCLC, in the task of classifying tumor versus normal tissue in Camelyon16, recall and precision have greater importance.

### 4.3 WSI Classification Results

To demonstrate the effectiveness of the attention-based feature distillation module proposed in this study in mitigating the interference of redundant features by constraining the selection of patches, thereby enhancing the classification performance of the model, the main experiments in this paper involve comparing the classification performance of multiple models on two publicly available datasets. In the classification tasks on both the Camelyon16 and NSCLC datasets, AFD-MIL achieved the best performance, as shown in Table 2. On the Camelyon16 dataset, the highest achieved ACC and AUC were 91.47% and 94.29%, respectively. On the NSCLC dataset, these two numbers were 93.33% and 98.17%. Compared to the original versions of ABMIL and TransMIL, AFD-MIL demonstrated significant improvements in performance (the green upward arrow in Table 2 indicates an improvement in performance, while the red downward arrow indicates a decline in performance). On the Camelyon16 dataset, AFD-MIL did not achieve the highest precision. This is because missing positive testing samples in this dataset poses a significant risk in the real world. Under the condition of maintaining a classifier decision threshold at 50%, the weights of the models that prioritize higher recalls were chosen as the final results. This decision reflects the importance of capturing positive cases, even at the cost of precision. Additionally, AFD-MIL outperformed the other model with a decoupled MIL module, IBMIL (SOTA), on both datasets. The attention-based feature distillation module designed in our study was able to extract more valuable features than the approach [35], while the global loss optimization addressed the issue of two-step MIL: it was hard to optimize the feature distillation module effectively. As a result, better classification performance has been achieved.

### 4.4 Comparison of Feature Distillation Methods

In this study, the number of features to be distilled, denoted as “$k$”, is a crucial parameter. Too few selected features can result in a significant loss of information during the feature distillation stage, and the model performance depends heavily on the training in this stage. On the other hand, too many selected features can increase interference in the attention network $f_{atten}$. Therefore, the choice of $k$ has a significant impact on the model performance. As indicated in Table 1, a WSI typically contains thousands of patches. In this study, we selected various values for $k$ to evaluate the performance differences resulting from changes in this parameter, as shown in Fig. 3. The experimental results indicate that as $k$ increases, the results of all the groups show an initial improvement followed by a decline. This result suggests that the number of features selected for feature distillation should be maintained at a certain level, such as $k = 16$ or $k = 32$. A too small $k$ value can result in too few features being passed to the subsequent network, making WSI classification challenging, while too large a $k$ value can lead to excessive features that may negatively impact the model performance. Simultaneously, as $k$ increases, the network handles more features, inevitably leading to computational burden. According to the results shown in Fig. 3, we have identified the convex point of the curve, and thus did not further increase $k$ to train the models.

### 4.5 Ablation Study

To validate the effectiveness of the proposed attention-based feature distillation and global loss optimization modules in this study, we conducted four sets of comparative experiments on both datasets by incrementally adding the newly designed components. The groups were as follows: 1. No feature distillation MIL (conventional ABMIL and TransMIL); 2. No attention-based feature distillation or global loss optimization, only instance-level feature distillation; 3. MIL with dual-channel feature distillation; and 4. Dual-channel feature distillation MIL and global loss optimization. As shown in Table 3, the experimental results indicate that both the new model structures and the optimization methods proposed in this study have led to a positive improvement in the classification task performance. AFD can improve the performance of the original two-step MIL model by optimizing the logic of feature distillation. This addresses the known issue of poor performance in the first step of the original two-step MIL, which is just based on instance-level MIL. Additionally, the introduction of global loss optimization changes the original situation, in which the feature

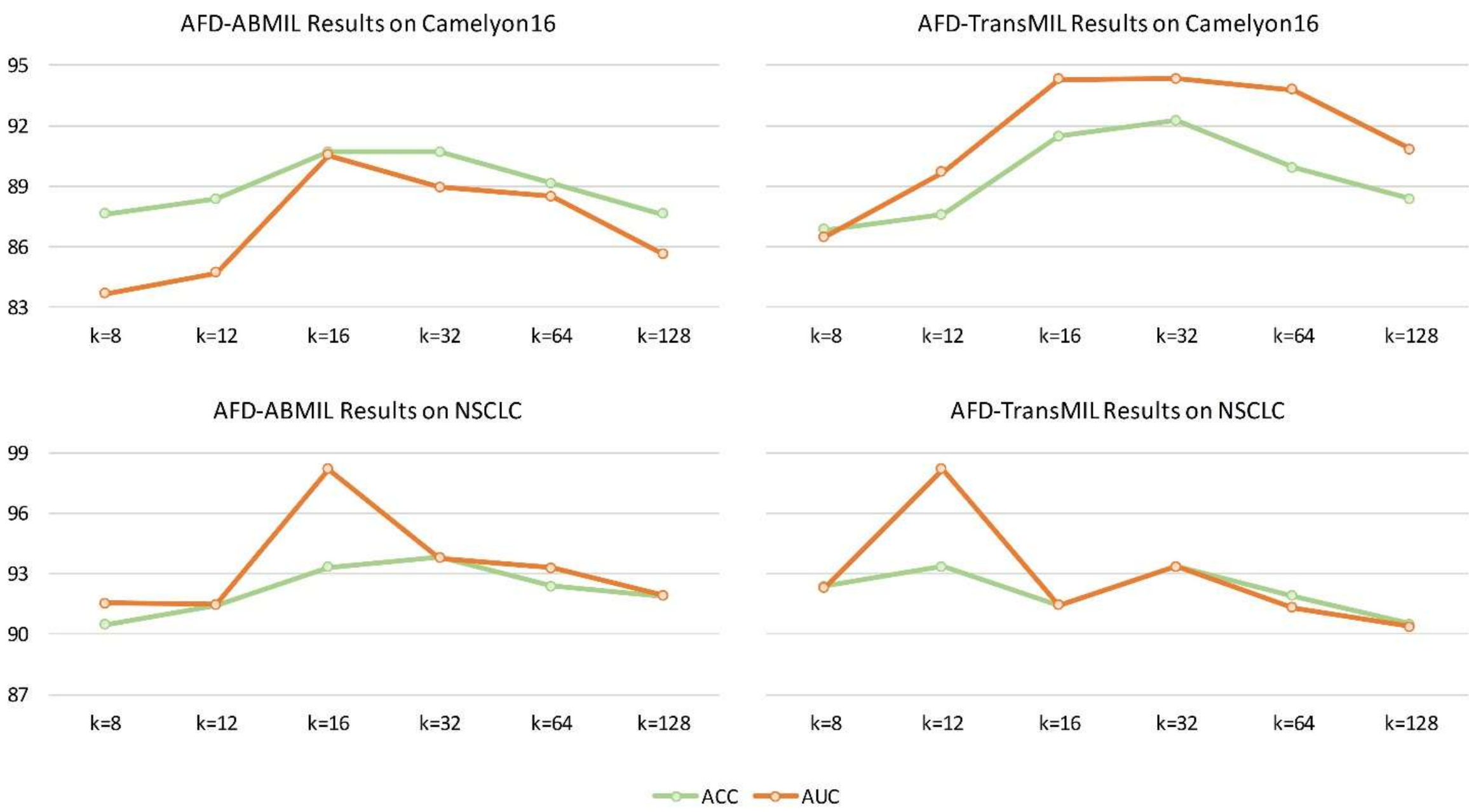

**Fig. 3.** The curve depicting the variation of model performance with the feature distillation parameter k.

TABLE 3

ABLATION STUDY RESULTS

| Component | | | ABMIL-based | | | | TransMIL-based | | | |
|---|---|---|---|---|---|---|---|---|---|---|
| | | | Camelyon16 | | TCGA-NSCLC | | Camelyon16 | | TCGA-NSCLC | |
| FD | Attention-based FD | Global Loss | ACC | AUC | ACC | AUC | ACC | AUC | ACC | AUC |
| × | × | × | 84.5 | 84.07 | 81.43 | 88.95 | 83.72 | 81.29 | 85.24 | 90.7 |
| √ | × | × | 82.95 | 82.77 | 81.9 | 88.91 | 88.37 | 93.62 | 90.95 | 97.88 |
| √ | √ | × | 88.37 | 86.28 | 92.38 | 92.32 | 89.92 | 93.78 | 92.38 | 97.15 |
| √ | √ | √ | 90.7 | 90.52 | 93.33 | 98.17 | 91.47 | 94.29 | 93.33 | 97.98 |

distillation's loss optimization is unrelated to the model's final prediction results, providing more fine-grained guidance for feature distillation.

### 4.6 Comparison of Feature Distillation Methods: Max-Positive and Max-Positive&Negative

As discussed in Section 3.a, Max-Positive actively selects the patches with the highest positive probability among all patches, which is more effective for the benign-malignant classification task on the Camelyon16 dataset. However, we hypothesize that this may not be the case for tasks involving classification of two subtypes of cancer (e.g., NSCLC). For negative samples, the patch with the highest probability of being positive may represent non-tumor tissue rather than the specific subtype of interest (representing negative tumors), such as lung adenocarcinoma. Therefore, we apply a maximum positive and negative method, which combines the patches most likely to be negative with those most likely to be positive. The results confirm our hypothesis. As shown in Table 4, the Max-Positive and Max-Positive&Negative methods achieved better performance on Camelyon16 and NSCLC, respectively. Although AFD-TransMIL with Max-Positive&Negative achieved higher accuracy on Camelyon16, the significant difference in recall compared to that achieved by Max-Positive is a substantial risk for the benign–malignant tumor classification task. Therefore, Max-Positive&Negative was not adopted for this dataset.

### 4.7 Heatmap Visualization of the Feature Distillation Module

To analyze the effectiveness of the feature distillation module in AFD-MIL for selecting patches, we compared the selected patches with the ground truth using heatmaps on the Camelyon16 test set, which contains pixel-level annotations. Fig. 4 depicts four sets of comparisons between the ground truth and the regions selected by AFD-MIL. It is evident that there is a substantial overlap between the ground truth (left, areas enclosed by the green curve while not encompassed by the red curve) and the regions selected by AFD-MIL (right, shaded in red, with the intensity of color indicating the probability of the patch being predicted as tumor by AFD-MIL). It is evident that there is a significant overlap between the ground truth and the regions selected by AFD-MIL. However,

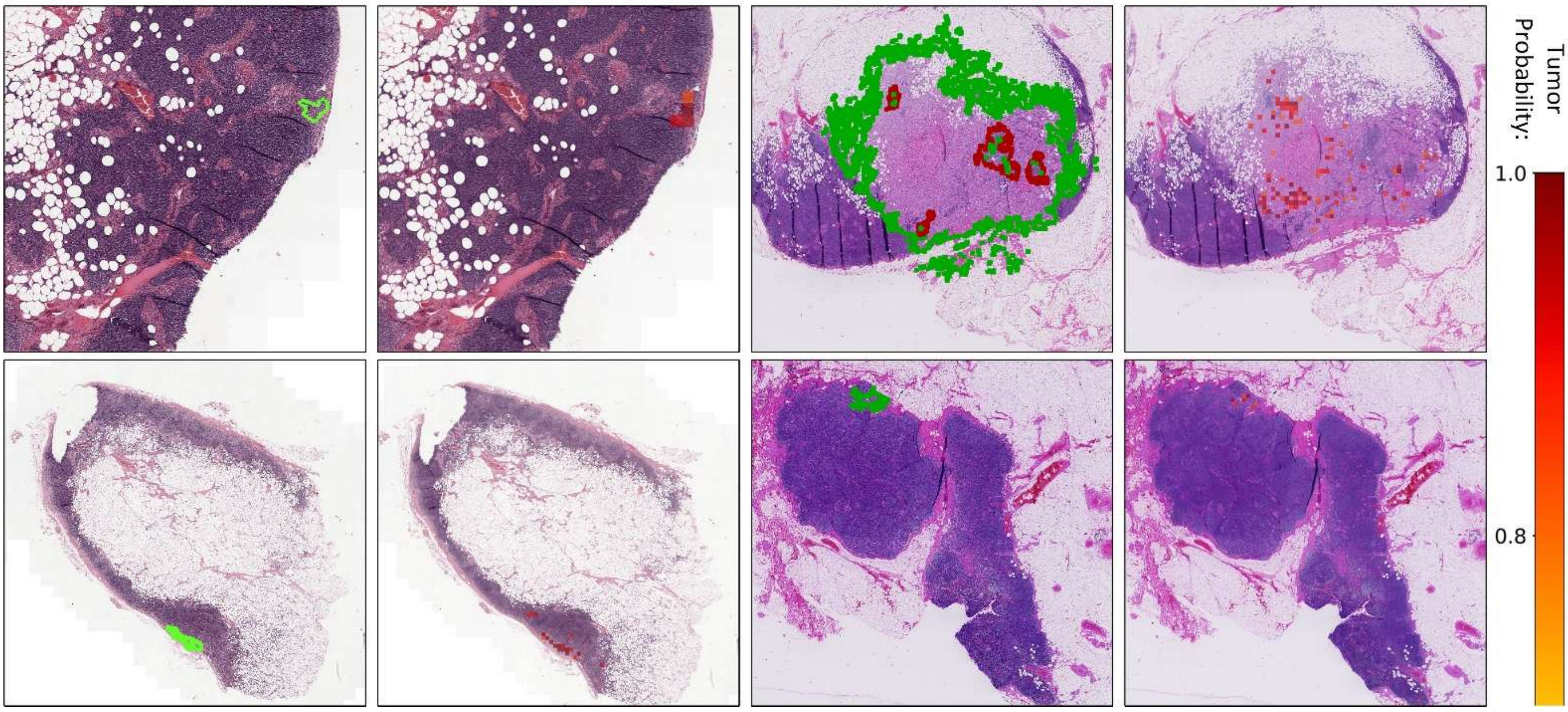

**Fig. 4.** The comparison heatmaps between the ground truth and the regions of interest identified by AFD-MIL.

TABLE 4

RESULTS OF DIFFERENT FEATURE DISTILLATION METHODS

| Methods | | Camlyon16 | | | | TCGA-NSCLC | | | |
|---|---|---|---|---|---|---|---|---|---|
| | | ACC | AUC | Recall | Precision | ACC | AUC | Recall | Precision |
| TransMIL | Max-Positive | 91.47 | 94.29 | 85.71 | 91.30 | 90.48 | 97.14 | 95.15 | 86.73 |
| | Max-Positive&Negative | 92.25 | 94.29 | 79.59 | 100.0 | 93.33 | 97.98 | 94.17 | 92.38 |
| ABMIL | Max-Positive | 90.70 | 90.52 | 89.8 | 86.27 | 89.52 | 0.8937 | 81.55 | 96.55 |
| | Max-Positive&Negative | 89.15 | 86.51 | 75.51 | 94.87 | 93.33 | 0.9817 | 93.26 | 93.64 |

the regions selected by AFD-MIL do not fully cover all ground truth regions. This is due to the relatively small parameter $k$. It is evident that AFD-MIL has already demonstrated the ability to select tumor regions similar to pathologists. AFD-MIL demonstrates the ability to accurately capture tumor features in pathological images, thereby enhancing the trustworthiness and credibility of its performance in classification tasks. This result demonstrates the potential application of this method in tasks such as tumor detection and diagnostic assistance, not limited to classification tasks. For example, leveraging AFD-MIL can provide cues to pathologists, assisting them in quickly and accurately completing diagnostic tasks.

## 5. CONCLUSION

The AFD-MIL proposed in this study is a novel orthogonal MIL methodology that is capable of significantly enhancing WSI classification performance. In contrast to past research that focused on optimizing image encoders and feature fusion mechanisms to eliminate redundant features, this study proposes a dual-channel feature distillation based on attention mechanisms and instance-level classification. This approach enables the identification of the most important patches in WSI, fundamentally eliminating redundant features. Additionally, compared to traditional two-step MIL, global loss optimization provides finer-grained guidance for the learning process of the feature distillation module. AFD-MIL achieves SOTA performance in both breast cancer and non-small cell lung cancer classification tasks. Furthermore, AFD-MIL exhibits excellent interpretability, with the selected patches highly overlapping with tumor regions. Due to its compatibility with other models, AFD-MIL has high scalability and is expected to be conveniently applied to clinical diagnostic assistance, providing fresh insight into MIL problem.

## 6. Acknowledgements

This work was supported in part by the China Key Research and Development Program (Grant No. 2023YFC2508200), the Fundamental Research Funds for the Central Universities (Grant No. N2219001), and the Liaoning Province Medical Engineering Cross Joint Fund (Grant No. 2022-YGJC-76).